\documentclass[final,5p,times,twocolumn]{elsarticle}
\usepackage{caption}
\usepackage{orcidlink}
\usepackage{academicons}
\usepackage{lineno,hyperref}
\usepackage{hyperref}
\hypersetup{
    colorlinks=true,
    linkcolor=blue,
    filecolor=magenta,
    urlcolor=blue,
    citecolor=blue,
    breaklinks=true,
}
\UseRawInputEncoding
\usepackage{amssymb}
\usepackage{CJK}
\usepackage{graphicx}  
\usepackage{dcolumn}   
\usepackage[latin1]{inputenc}
\usepackage[T1]{fontenc}
\usepackage[autolanguage]{numprint} 
\usepackage{hyphenat}
\usepackage{amsmath}
\usepackage{lscape}
\usepackage{booktabs,longtable}
\usepackage{mathptmx, courier, pifont}
\usepackage[scaled=0.92]{helvet}
\usepackage{textcomp}
\usepackage{rotating}
\usepackage{verbatim}
\usepackage{picinpar}
\usepackage{wrapfig}
\usepackage{float}
\usepackage{multirow}
\usepackage{xcolor}
\usepackage[normalem]{ulem}
\definecolor{orcidlogocol}{HTML}{A6CE39}

\makeatletter
\def\ps@pprintTitle{%
 \let\@oddhead\@empty
 \let\@evenhead\@empty
 \def\@oddfoot{\footnotesize\itshape
       Published in \textit{Nucl. Instrum. Methods Phys. Res. A}, 1086 (2026) 171337, DOI: https://doi.org/10.1016/j.nima.2026.171337 \hfill \thepage}%
 \let\@evenfoot\@oddfoot}
\makeatother

\begin{document}
\begin{frontmatter}

\title{Data sorting modes of phoswich detector array}

\author[ijclab]{R.~Li\corref{cor1}\orcidlink{0000-0002-2782-2333}}
\ead{liren824@gmail.com}
\cortext[cor1]{Corresponding author.}
\author[ijclab]{D.~Verney\orcidlink{0000-0001-7924-2851}}
\author[ijclab,jyvaskyla]{C.~Delafosse\orcidlink{0000-0001-5717-2426}}
\author[KVI]{M. N.~Harakeh\orcidlink{0000-0002-7271-1712}}
\author[krakow]{A. ~Maj\orcidlink{0000-0002-2873-7077}}
\author[iphc]{F.~Didierjean\orcidlink{0009-0002-4950-3162}}

\author[ijclab,jyvaskyla]{L.~Al~Ayoubi\orcidlink{0000-0002-1441-9094}}
\author[lebanon]{H.~Al Falou\orcidlink{0009-0002-0516-498X}}
\author[krakow]{P. ~Bednarczyk\orcidlink{0000-0002-5699-5292}}
\author[infn]{G.~Benzoni\orcidlink{0000-0002-7938-0338}}
\author[ijclab]{F.~Le Blanc}
\author[nigde]{V.~Bozkurt\orcidlink{0000-0003-4651-0447}}
\author[krakow]{M.~Ciema\l{}a\orcidlink{0000-0001-5328-720X}}
\author[infn,milanoU]{F. C. L. Crespi\orcidlink{0000-0003-2254-4255}}
\author[ijclab]{I.~Deloncle}
\author[ijclab]{C.~Gaulard}
\author[lnl]{A.~Gottardo\orcidlink{0000-0002-0390-5767}}
\author[subatech]{V.~Guadilla\fnref{label1}\orcidlink{0000-0002-9736-2491}}
\author[ijclab]{J.~Guillot\orcidlink{0000-0001-9650-0613}}
\author[warsaw]{K.~Hady\'{n}ska-Kl\k{e}k}
\author[ijclab]{F.~Ibrahim}
\author[ijclab]{N.~Jovancevic\fnref{label2}}
\author[jyvaskyla]{A.~Kankainen\orcidlink{0000-0003-1082-7602}}
\author[krakow]{M. ~Kmiecik\orcidlink{0000-0003-2019-2399}}
\author[ijclab]{M.~Lebois}
\author[ciemat]{T.~Mart\'{i}nez\orcidlink{0000-0002-0683-5506}}
\author[warsaw]{P.~Napiorkowski}
\author[ijclab]{B.~Roussiere}
\author[jinr]{Yu.~G.~Sobolev\orcidlink{0000-0002-5615-0468}}
\author[horia]{M. ~Stanoiu\orcidlink{0000-0002-9992-9463}}
\author[ijclab]{I.~Stefan\orcidlink{0000-0001-7923-8908}}
\author[jinr]{S.~Stukalov\orcidlink{0000-0003-2948-1947}}
\author[ijclab]{D.~Thisse}
\author[ijclab]{G.~Tocabens}

\address[ijclab]{Universit\'e Paris-Saclay, CNRS/IN2P3, IJCLab, 91405 Orsay, France}
\address[jyvaskyla]{University of Jyv\"askyl\"a, Department of Physics, Accelerator Laboratory, P.O. Box 35, FI-40014, Finland}
\address[KVI]{ESRIG, University of Groningen, Zernikelaan 25, 9747 AA Groningen, The Netherlands}
\address[krakow]{Institute of Nuclear Physics, Polish Academy of Sciences, Krakow, Poland}
\address[iphc]{Institut Pluridisciplinaire Hubert Curien, CNRS/IN2P3 and Universit\'e de Strasbourg, Strasbourg, France}
\address[lebanon]{Faculty of Sciences 3, Lebanese University, Michel Slayman Tripoli Campus, Ras Maska 1352, Lebanon}
\address[infn]{INFN, sezione di Milano, Dipartimento di Fisica, Milano, Italy}
\address[nigde]{Department of Physics, Science Faculty, Nigde Omer Halisdemir University, 51240 Nigde, Turkey}
\address[milanoU]{Dipartimento di Fisica dell\textquoteright{}Universit\`a degli Studi di Milano, I-20133 Milano, Italy}
\address[lnl]{Laboratori Nazionali di Legnaro, I-35020 Legnaro, Italy}
\address[subatech]{Subatech, CNRS/IN2P3, Nantes, EMN, F-44307, Nantes, France}
\address[ciemat]{Centro de Investigaciones Energ{\'e}ticas, Medioambientales y Tecnol{\'o}gicas (CIEMAT), Madrid, Spain}
\address[warsaw]{Heavy Ion Laboratory, University of Warsaw, 02-093 Warsaw, Poland}
\address[jinr]{Joint Institute for Nuclear Research, Dubna, Russia}
\address[horia]{Horia Hulubei National Institute of Physics and Nuclear Engineering, 30 Reactorului, 077125 Bucharest,Romania}
\fntext[label1]{Present address: Faculty of Physics, University of Warsaw, 02-093 Warsaw, Poland.}
\fntext[label2]{Present address: University of Novi Sad, Faculty of Science, Novi Sad, Serbia.}

\begin{abstract}

The different data-sorting modes of the phoswich detector array PARIS used for detecting high-energy (4$-$10 MeV) $\gamma$ rays are investigated. The characteristics including time resolution, energy resolution and detection efficiency under various modes are studied. The present study shows that PARIS has capabilities of rejecting escape and pileup events when used for decay spectroscopy. Notably, the methods presented in this work refer specifically to the $\beta$-decay experiment of $^{80g+m}$Ga conducted with three PARIS clusters comprising 27 phoswich detectors, rather than to a general report on the PARIS array or its overall performance for in-beam spectroscopy. Compared with the 2"$\times$2"$\times$2" LaBr$_3$(Ce) detector (Ciema{\l}a et al., 2009), even in individual mode, PARIS provides significant suppression of single- and double-escape peaks and reduces background via vetoing function of the outer-volume NaI(Tl) crystals. In contrast to the common approach of adding back the energies in LaBr$_3$(Ce) and NaI(Tl) to increase the detection efficiency of the full-energy peak, using NaI(Tl) as a veto shield provides a superior trade-off for applications where spectral purity is essential. Employing add-back analysis within each cluster of nine phoswiches or between all phoswiches could enhance full-energy peak efficiency and further suppress escape peaks and background. Applying a multiplicity condition provides a further suppression but simultaneously lowers the statistics of full-energy peaks. 

\end{abstract}


\end{frontmatter}

\section{Introduction}
For the purpose of measuring high-energy $\gamma$ rays, e.g., 4$-$10 MeV, a novel 4$\pi$ calorimeter PARIS (Photon Array for studies with Radioactive Ion and Stable beams) was designed \cite{roberts2011testing,Ghosh2016,Wasilewska_2015}. These $\gamma$ rays could be emitted from nuclear bound states located close to particle separation energy, e.g., pygmy dipole resonance (PDR) \cite{bracco2019isoscalar} states, and also could be from neutron/proton-unbound states, e.g., isovector giant dipole resonance states (IVGDR) \cite{harakeh2001giant}, etc. These spectroscopic studies are helpful for understanding the nuclear structure in the bound and continuum regions including collectivity, interplay between collective motion and single-particle excitation and have important applications in astrophysics, e.g., investigating the competition between neutron- and $\gamma$-emissions, excited-states-based PDR \cite{li_tel,PhysRevC.110.064323}, impacts of nuclear structure on $\beta$ decay \cite{PhysRevC.111.034303,xtvt-qy6v}, validity of Brink-Axel Hypothesis \cite{brink1955some,axel1962electric} in the high-excitation energy region, etc.

In order to guarantee detection efficiency, that is critical for compensating the beam intensity limit of exotic radioactive ions, several PARIS clusters are usually used to constitute a versatile array. Each cluster is comprised of 9 optically isolated segmented phoswiches \cite{PARIS}. A phoswich is composed of two different scintillators. The inner volume (that is, situated closer to the target when in use) is a 2"$\times$2"$\times$2" Lanthanum Bromide/Cerium Bromide (LaBr$_3$/CeBr$_3$) crystal providing high detection efficiency, excellent time resolution and relatively good energy resolution in a large energy range. The outer volume (situated directly behind the inner LaBr$_3(Ce)$) is a more conventional scintillator, 2"$\times$2"$\times$6" Thallium-doped Sodium Iodide (NaI(Tl)). Note that the two volumes are optically connected and jointly enclosed in an aluminum cover. The light outputs generated in both scintillators are collected by a common 8 stage, 46 mm diameter Hamamatsu photomultiplier tube (PMT) R7723-100 at the rear end \cite{hamamatsu}. Compared to high-purity germanium (HPGe) detectors, LaBr$_3$(Ce) has several unique characteristics: 1) no cryogenic cooling is needed, making it convenient to use; 2) high intrinsic detection efficiency; 3) fast time response. Furthermore, PARIS has high granularity that allows to measure discrete transitions and the multiplicities of cascades which are crucial for the assignments of spins and parities to nuclear states. An extremely interesting property of PARIS is that the outer-volume NaI(Tl) can play the role of veto detector to perform anti-coincidence operation for rejecting the escape events. Additionally, PARIS can reject pileup events through the technique of Pulse Shape Analysis (PSA). These properties are significant for enhancing peak-to-background ratio. Regarding data analysis, given the complex configuration, PARIS has different data-sorting modes.  

\begin{figure}[!htb]
\centering
\includegraphics[width=0.8\columnwidth]{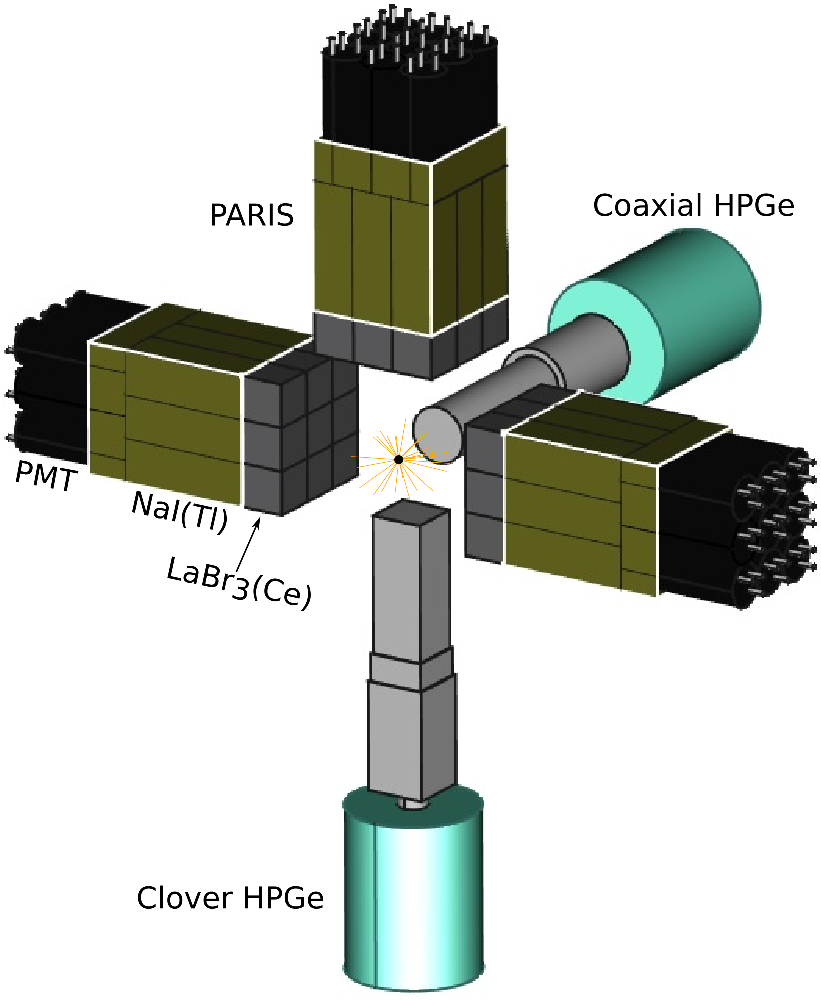}
\caption{Schematic drawing of hybrid $\gamma$-ray spectrometer. Three PARIS clusters are shown, each comprising 9 phoswiches with the two-inch cubes of LaBr$_3$(Ce) backed by the six-inch long NaI(Tl) and then the photomultiplier tube. The distances between the source collection point and phoswich detectors of PARIS, clover HPGe and coaxial HPGe are 120, 74, and 50 mm, respectively.}
\label{fig0_0}
\end{figure}

In this article, along with the energy resolutions for one of the detectors and for one entire cluster, we (1) provide the first evaluation of the online performance of the PARIS array composed of 3 clusters (27 phoswiches) as shown in Fig. \ref{fig0_0}, e.g., the time resolution, the energy resolution, and detection efficiencies; (2) present various data-sorting modes tailored for the phoswich array, from each phoswich being an individual detector to TAS-like (Total Absorption Spectroscopy) calorimeter; (3) demonstrate the background suppression capabilities of PARIS using NaI(Tl) as a veto, which offers a superior trade-off for applications where spectral purity is essential. The background-free spectra shown are a direct result of this approach. As a continuation of the detector development, this work builds upon Ref. \cite{Ghosh2016}, which reported the performance of two individual phoswich detectors including the reconstruction of total energy in phoswich components using NaI(Tl) as part of the energy add-back. It also builds upon Ref. \cite{Wasilewska_2015}, which presented $\gamma$ spectra up to 8.9 MeV with different multiplicities and with external add-back (within one cluster).

\section{Characterization of phoswich detector array}
\subsection{Signal separation} \label{signal-separation}
The first step of processing phoswich detector array PARIS data is to separate the signals coming from LaBr$_3$(Ce) and NaI(Tl) crystals, which is prerequisite for their individual energy and detection efficiency calibrations and time alignment. The PSA method is used to resolve the two signals from one phoswich using their widely different shaping and decay times as shown by Fig. 2 in Ref. \cite{Ghosh2016}. So, one saves two values digitized by a triggerless digital data acquisition system (DAQ) called FASTER (Fast Acquisition System for nuclEar Research) developed at the Laboratoire de Physique Corpusculaire (LPC) in Caen \cite{FASTER}: the signal recorded by the QDC is integrated over a shorter time gate (120 ns) and a longer time gate (820 ns) matched to the first part and the decaying tail of the signal. The short time gate is to cover full rise and decay of the LaBr$_3$(Ce) signal while the long time gate is determined by the NaI(Tl) signal decay time.

\begin{table*}[htbp]
\centering
\fontsize{10pt}{11pt}\selectfont 
\caption{Energy deposition scenarios when a $\gamma$ ray hits one phoswich.}  \vspace{0.3mm}
\label{tab1}
\renewcommand\arraystretch{1.2}
\begin{tabular}{ll}
\hline
\hline
Situations & Description \\
\hline
A & Deposit full energy in one LaBr$_3$ crystal	\\
B & Deposit part energy in one LaBr$_3$ crystal and the rest in neighboring LaBr$_3$(s) within one cluster\\
C & Deposit part energy in one LaBr$_3$ crystal and the rest in other LaBr$_3$(s) inter-cluster\\
D & Deposit full energy in one NaI(Tl) crystal \\
E & Deposit part energy in one LaBr$_3$ crystal and the rest in the outer-volume NaI(Tl)(s)\\
F & Deposit part energy in one LaBr$_3$ crystal and part in outer-volume NaI(Tl)(s)\\
G & More than one $\gamma$ ray hit the same phoswich within the long QDC window of 820 ns, pile-up event\\
\hline
\hline
\end{tabular}
\end{table*}

\begin{figure}[!htb]
\centering
\includegraphics[width=1.0\columnwidth]{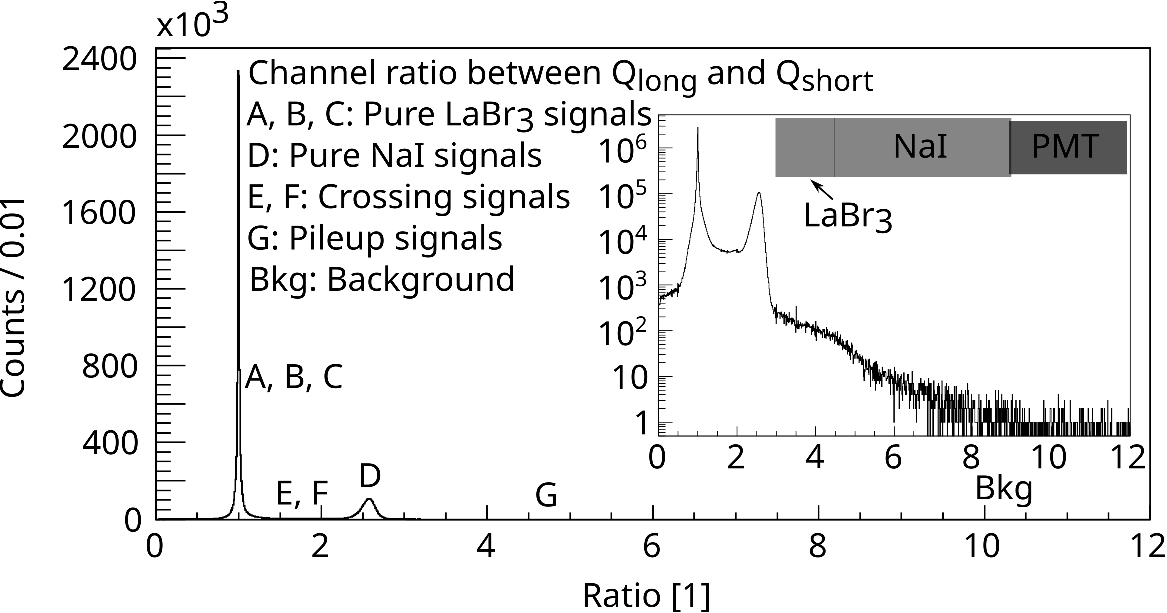}
\caption{Ratio of output signals between Q$_{long}$ and Q$_{short}$, Q$_{long}$/Q$_{short}$, from one phoswich digitized by FASTER QDC. The inset shows the same spectrum on a logarithmic scale and a schematic drawing of one phoswich detector.}
\label{fig1}
\end{figure}

There are seven different situations when a $\gamma$ ray arrives into a phoswich as listed in Tab. \ref{tab1}: A. the $\gamma$ ray hits the LaBr$_3$ crystal and deposits its full energy inside. In this situation, one observes the same value in Q$_{short}$, collected charge in PMT with shorter QDC time gate of 120 ns, and Q$_{long}$ with a longer time gate of 820 ns; see an example of a two-dimensional Q$_{short}$-Q$_{long}$ plot in Fig. 2 of Ref. \cite{Wasilewska_2015}. Hence, the ratio between Q$_{long}$ and Q$_{short}$, Q$_{long}$/Q$_{short}$, is equal to unity as shown by the first separated peak in Fig. $\ref{fig1}$. B. same situation as 1 but a few channels are fired within one cluster. C. same situation as 1 but a couple of channels are fired in different clusters. D. $\gamma$ ray goes through LaBr$_3$ but deposits zero energy and then hits NaI(Tl), depositing full or partial energy there. This gives rise to another clear peak in the spectrum of Fig. $\ref{fig1}$. The ratio between Q$_{long}$ and Q$_{short}$ is larger than 2 because for pure NaI(Tl) signal the collected charges are smaller than half of full charges in a 120 ns time gate. E. $\gamma$ ray punches through LaBr$3$ and NaI(Tl), depositing part energy in LaBr$_3$(Ce) and the rest in NaI(Tl). These events are localized in the crossing part in Fig. $\ref{fig1}$. Since a $\gamma$ ray deposits partial energy in LaBr$_3$(Ce) and the rest in NaI(Tl), the ratio is larger than the first peak but smaller than the second peak in Fig. $\ref{fig1}$. F. Same as E but a $\gamma$ ray deposits partial energy in NaI(Tl). Then the photon escapes the phoswich from the bottom or sides. These events are localized in the crossing part in Fig. $\ref{fig1}$ as well. G. If more than two $\gamma$ rays hit the phoswich at the same time (pileup events), the time difference between arrivals is smaller than 820 ns. These events are visible in the region labeled "G" in Fig. \ref{fig1}. Additionally, the background is distinguishable as well, shown in the region labeled "Bkg" with a ratio between 9 and 12. We assign this region to background because the ratio is too large and the spectral structure differs from that of pileup events. The statistics in this region exhibit random distribution with values between 0 and 2. Given the beam intensity of around 10$^4$, the ability of the FASTER card to encode up to 500 million events per second per channel, the good time resolution of the detectors, and the typically small number of $\gamma$ rays in the cascade populated in the $\beta$-decay reaction, we do not have serious pileup events with ratios larger than 9. Notably, Tab. \ref{tab1} does not account for the scenario where $\gamma$ rays pass through LaBr$_3$(Ce) without depositing energy and subsequently interact with more NaI(Tl) detectors, either within the same cluster or across different clusters. That is because these events are limited and appear in the low-energy region. A single NaI(Tl) is 6-inch long and therefore it can cover Compton-scattered escaping events with angles of 0$^{\circ}$ - 56.3$^{\circ}$ when a photon hits the center of the first 2-inch section of the crystal. According to the Klein-Nishina formula, for a $\gamma$ ray with energy larger than 2 MeV, the scattering cross section for an angle larger than 56.3$^{\circ}$ is rather small. Furthermore, photons scattered at large angles are mainly in the 0 - 2 MeV energy range. Usually, we use PARIS to measure high-energy $\gamma$ rays taking advantage of its high detection efficiency. Notably, without applying anti-coincidence analysis to suppress escape events using the outer-volume NaI(Tl), the peaks in the high-energy region, e.g., 4$-$10 MeV, will be difficult to observe. Therefore, with add-back between NaI(Tl) crystals one cannot obtain a background-free spectrum. Firstly, one can remove the bottom-escape events from the spectra by using NaI(Tl) as veto detectors and pileup events by selecting the proper region in Fig. $\ref{fig1}$. Secondly, the photon can escape the phoswich from the side and then hit a neighboring phoswich in the same cluster or in another cluster. These side escape events can be suppressed by applying add-back procedure either within a single cluster or globally across all phoswich detectors. It is important to note that the presence of an aluminum layer between the phoswich detectors may affect the efficiency of the process to some extent.

After the above analysis, events with full energy deposited in LaBr$_3$(Ce) and events with full/partial energy deposited in NaI(Tl) could be distinguished using Fig. \ref{fig1}. Furthermore, one can generally convert measured values of Q$_{short}$ and Q$_{long}$ to quantities q$_1$(E$_{LaBr3}$(Ce)) and q$_2$(E$_{NaI(Tl)}$) which contain the LaBr$_3$(Ce) and NaI(Tl) energy deposit information exclusively. Then one gets E$_{LaBr_3(Ce)}$ and E$_{NaI(Tl)}$ by calibrating q$_1$(E$_{LaBr_3(Ce)}$) and q$_2$(E$_{NaI(Tl)}$) using reference $\gamma$ sources; see more details in Fig. 3.13 of Ref. \cite{li_tel} and the related description therein. This process is called the signal separation procedure. The procedure as described so far allows the energy deposited in each part of the phoswich to be calibrated and recorded.

\begin{figure}[!htb]
\centering
\includegraphics[width=0.9\columnwidth]{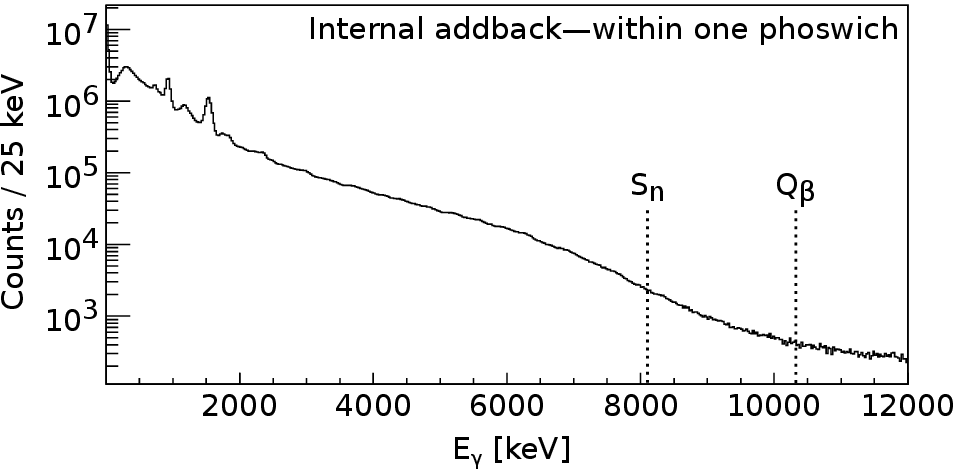}
\caption{$\beta$-gated $\gamma$-ray spectrum of $^{80}$Ge from PARIS, where energy signals from LaBr$_3$(Ce) and NaI(Tl) components of each phoswich detector are added together (internal add-back).}
\label{fig1_1}
\end{figure}

The $\gamma$ spectrum originating solely from LaBr$_3$(Ce) is then obtained by selecting events where E$_{NaI(Tl)}$ = 0. Using the same method, by gating on events with E$_{LaBr_3(Ce)}$ = 0, one gets the spectrum exclusively from the NaI(Tl) crystal as well. The inner-outer-volumes add-back spectrum is obtained by adding the deposited energies in LaBr$_3$(Ce) and NaI(Tl) crystals, i.e., E$_{LaBr_3(Ce)}$ + E$_{NaI(Tl)}$. However, it is worth pointing out that the energy resolution of the $\gamma$-spectrum will deteriorate due to events not depositing their full energy in the NaI(Tl) crystals, though one obtains higher statistics when applying this procedure. Fig. \ref{fig1_1} shows the $\beta$-delayed $\gamma$-spectrum of $^{80}$Ge using this method.

\subsection{Time resolution}

\begin{figure}[!htb]
\centering
\includegraphics[width=1.0\columnwidth]{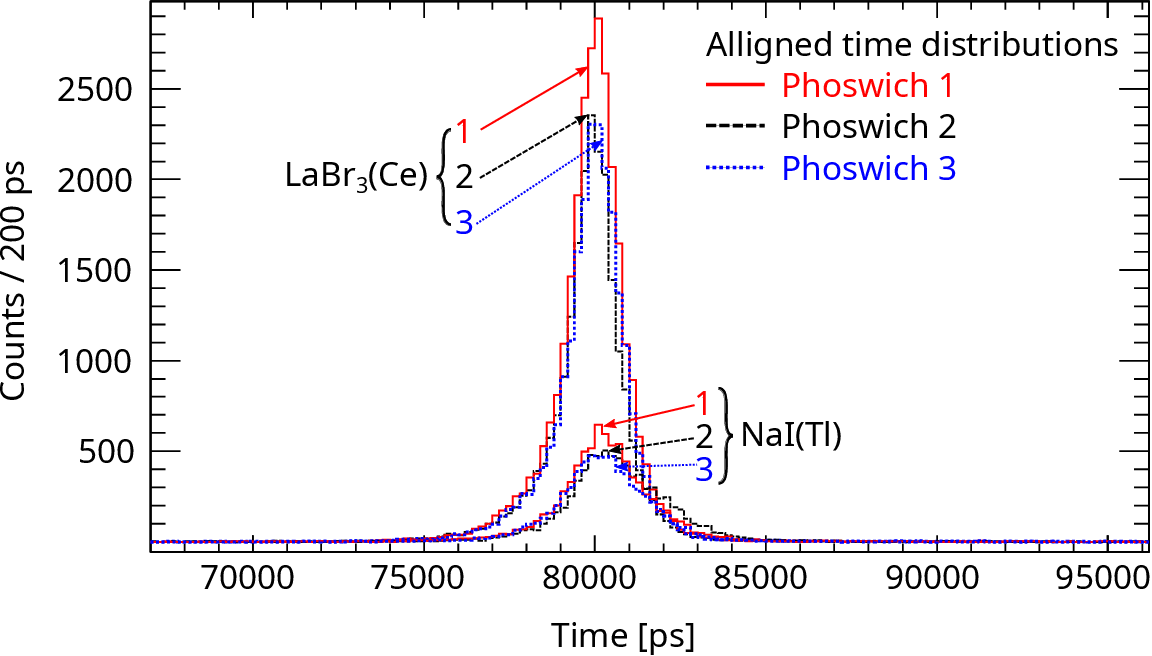}
\caption{Time-signal distributions of NaI(Tl) (the peaks with the lowest counts) and LaBr$_3$(Ce) (the peaks with the highest counts) channels relative to $\beta$-particle signal measured by a cylindrical plastic scintillator surrounding the source collection point with a time resolution of ps magnitude, arbitrarily aligned to the 80 ns position. 1, 2 and 3 are three separate phoswiches, represented by the solid (red), dashed (black) and dotted (blue) lines, respectively, in the figure. Note that $\gamma$ rays of the cascade of the isomer should be avoided when performing time alignment. The typical time resolution observed for LaBr$_3$(Ce) and NaI(Tl) is listed in Tab. \ref{tab2}.}
\label{fig2}
\end{figure}

It is essential in the data processing for the energies recorded in each detector for genuine coincidences to be assigned the same event time. The timing information in the present experiment was extracted via digital constant-fraction discrimination (CFD) using the FASTER DAQ system. However, delays may appear between different channels due to several reasons like different signal rising time, different cable lengths, detector jitter, electronic jitter, DAQ jitter including sampling and clock instabilities, temperature changes, electromagnetic interference, mechanical vibration, etc. Fig. $\ref{fig2}$ presents the result of time alignment of labels 17, 18, 19 corresponding to three phoswiches of PARIS. They were aligned to 80 ns arbitrarily using the online ion source of $^{80g+m}$Ga, based on the time difference between the $\beta$ particle detected in the plastic scintillator and the selected $\gamma$ ray recorded in the LaBr$_3$(Ce) or NaI(Tl) scintillator. Note that $\gamma$ rays of the cascade of the isomer should be avoided when performing time alignment. For each phoswich the times of LaBr$_3$ and NaI(Tl) events were corrected separately. Note that the quality of time correction is vital for fast timing analysis using LaBr$_3$(Ce) to extract the lifetime of the state. 

\begin{figure}[!htb]
\centering
\includegraphics[width=1.0\columnwidth]{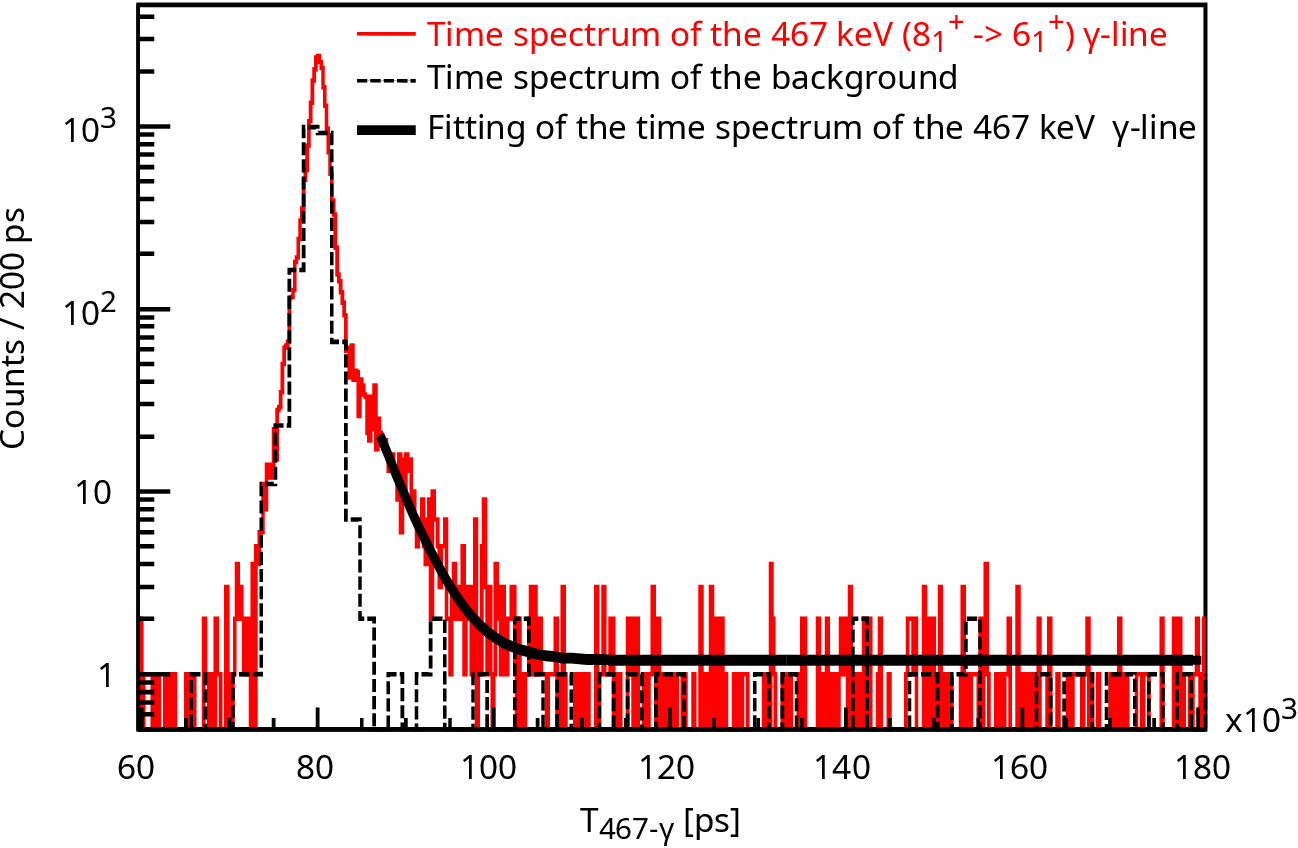}
\caption{Time spectrum of the 467 keV $\gamma$-line (8$_1^+$ $\rightarrow$ 6$_1^+$ transition in $^{80}$Ge) and its fitting. The black dashed curve is the time spectrum of background $\gamma$ rays adjacent to the 467 keV $\gamma$-line, which is plotted as a reference. In these spectra, the time resolution of a LaBr$_3$(Ce) crystal in PARIS and the delayed-tail of the 467 keV $\gamma$-line are clearly distinguishable. The extracted half-life of the 3445.3(6) keV isomeric state, 8$_1^+$, of $^{80}$Ge in the present work is 3.08(6) ns.}
\label{fig2_1}
\end{figure}

One advantage of the PARIS detector is its excellent time resolution, which originates from the fast responses of LaBr$_3$(Ce) crystals and FASTER DAQ. It gives to PARIS the capability of fast-timing measurement in addition to other functions. The design objective of time resolution ($\sigma$) of PARIS at 1 MeV is 250 ps, which makes many lifetime measurements possible for nuclear excited states with half-lives around or longer than 1 ns using easy data analysis method, and also allows the use of time-of-flight (TOF) techniques to distinguish between $\gamma$-ray and neutron energy deposits in the crystals. A cylindrical plastic scintillator surrounding the collection point used for $\beta$-tagging provides a beginning time for the TOF measurement. Furthermore, given the excellent time performance of the LaBr$_3$(Ce) crystal, the FWHM of the time resolution is extracted using the time difference between two cascade $\gamma$ rays (659.2 and 1083.6 keV), detected by two phoswich detectors, in order to avoid the influence of the plastic detector. The time resolution of PARIS (LaBr$_3$(Ce) crystal) in the $^{80}$Ga $\beta$-decay experiment is 437(49) ps as listed in Table $\ref{tab2}$. The time resolutions of NaI(Tl) crystal in PARIS, coaxial HPGe and clover HPGe are also tabulated, which is important information for the determination of the time window of $\gamma$-$\gamma$ coincidence. Note that the time resolution of clover HPGe detector is derived from the summed spectrum of its four crystals using the TOF method. The time resolutions of NaI(Tl) are determined from the summed spectra of 27 phoswich detectors using the TOF method. The time spectra used represent the time differences between $\beta$ particles detected by the plastic scintillator and the subsequent 1083 keV $\gamma$ rays recorded in LaBr$_3$(Ce), NaI(Tl), coaxial HPGe, and clover HPGe detectors. The extracted half-life of the 3445.3(6) keV isomeric state, 8$_1^+$, of $^{80}$Ge in the present work is 3.08(6$_{stat}$) ns via fitting the time spectrum of its deexcitation 466.7(3) keV (8$_1^+$ $\rightarrow$ 6$_1^+$) $\gamma$-line as shown in Fig. \ref{fig2_1}, that is in good agreement with previous measurements with BaF$_2$ scintillators \cite{Mach_2005}.

\begin{table}[!htb]
\centering
\fontsize{8.55pt}{11pt}\selectfont
\caption{Time resolution (full width at half maximum) of the various detectors at 1083 keV in units of ns. Note that the time resolution of the clover HPGe detector is derived from the summed spectrum of its four crystals using the TOF method.} 
\vspace{0.2mm}
\label{tab2}
\renewcommand\arraystretch{1.5}
\begin{tabular}{lllll}
\hline
\hline
Detectors	& LaBr$_3$(Ce) & NaI(Tl) & Coaxial HPGe	& Clover HPGe \\
\hline
FWHM	& 0.437(49) & 2.78(4)	& 10.39(2)	& 12.65(5)	\\
\hline
\hline
\end{tabular}
\end{table}

\subsection{Energy resolution}

\begin{figure}[!htb]
\centering
\includegraphics[width=1.0\columnwidth]{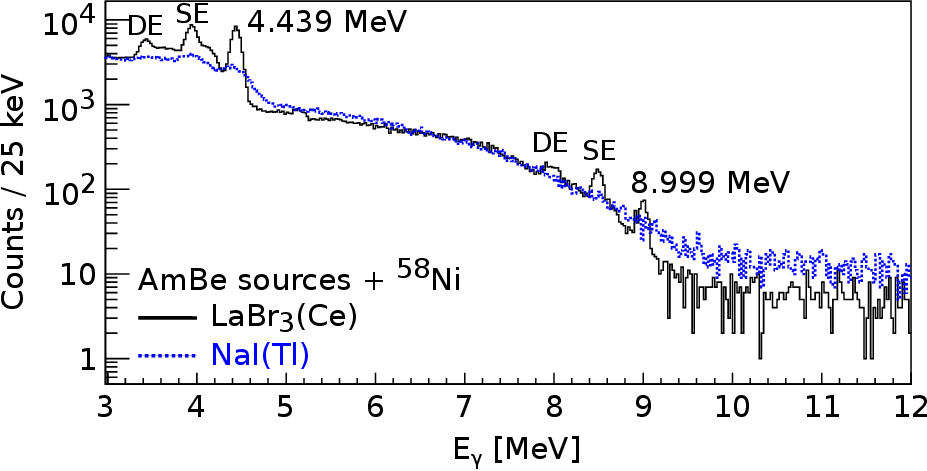}
\caption{$\gamma$-spectra emitted by $^{12}$C and $^{59}$Ni, created in the $^9$Be($\alpha$,n$\gamma$) and $^{58}$Ni(thermal neutron capture,$\gamma$) reactions, respectively, with $\gamma$-rays of 4438.94 keV and 8998.63 keV energies, measured by a phoswich detector of PARIS composed of a 2"$\times$2"$\times$2" LaBr$_3$(Ce) (black curve) and a 2"$\times$2"$\times$6" NaI(Tl) (dotted blue) scintillators. SE and DE are abbreviations of single-escape and double-escape, respectively. Note that the two spectra were acquired simultaneously.}
\label{fig5_0}
\end{figure}

\begin{figure}[!htb]
\centering
\includegraphics[width=1.0\columnwidth]{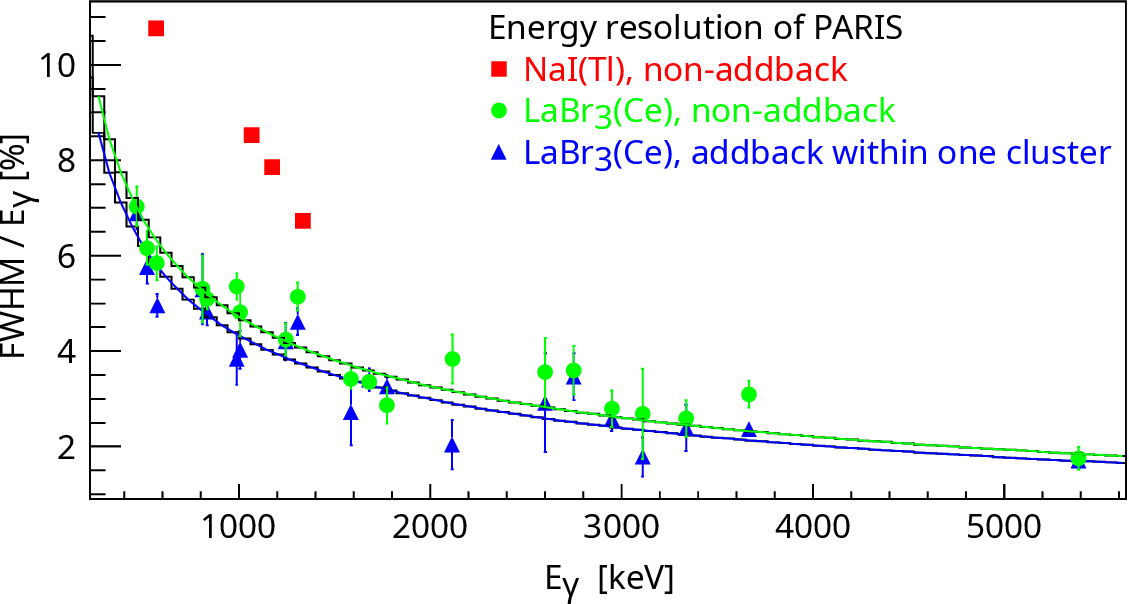}
\caption{Energy resolution as a function of $\gamma$-ray energy of the LaBr$_3$(Ce) shown as green circles and NaI(Tl) components shown as red squares of the 27 PARIS phoswich detectors, obtained from the non-add-back spectra. The blue triangles represent the energy resolution of the LaBr$_3$(Ce) component from the add-back within-one-cluster spectrum. Fitting function used is FWHM/E$_{\gamma}$ = a(E$_{\gamma}$+b*E$_{\gamma}^2$)$^{-1/2}$. The data used are tabulated in Tab. \ref{tab2_1}. Note that the smaller uncertainties in the data obtained with NaI(Tl) are due to higher statistics collected with $^{207}$Bi and $^{60}$Co sources \cite{gamma-source}, compared to those obtained with LaBr$_3$(Ce) using $\beta$-delayed $\gamma$ rays of $^{80}$Ge \cite{PhysRevC.111.034303}}
\label{fig5}
\end{figure}

Given the complex structure of the phoswiches and the response of the digitization electronics, the energy calibration of PARIS is decomposed into four steps: 1. distinguish between LaBr$_3$(Ce) and NaI(Tl) signals; 2. calibrate LaBr$_3$ and NaI(Tl) detectors separately for low-energy parts, below 4.5 MeV, using standard $\gamma$ sources including $^{137}$Cs for 661.66 keV, $^{60}$Co for 1173.23 keV, 1332.50 keV and 2505.69 keV, $^{207}$Bi for 569.70 keV, 1063.66 keV and 1770.23 keV, AmBe for 4438.94 keV and some $\gamma$ rays from background like $^2$H for 2223.25 keV populated by hydrogen neutron capture, $^{214}$Bi for 2204.10 keV and $^{208}$Tl for 2614.51 keV; 3. correct low-energy part calibration using well identified $\gamma$ rays from the on-line ion source; 4. calibrate LaBr$_3$(Ce) and NaI(Tl) crystals separately for high-energy part, above 4.5 MeV, using AmBe for 4438.94 keV and AmBe + Ni for 8998.63 keV issued from $^{59}$Ni after neutron capture reaction \cite{603792}. Fig. \ref{fig5_0} presents the $\gamma$-spectra of an AmBe source surrounded by $^{58}$Ni metal measured by a phoswich of PARIS. After calibration, the energy resolution of the phoswich detector array as a function of $\gamma$-ray energy is obtained. The FWHM/E values for LaBr$_3$(Ce) and NaI(Tl), obtained with 27 phoswich detectors in non-add-back mode, are 4.8(5)$\%$ at 1005 keV and 7.86(6)$\%$ at 1173 keV, respectively, as shown in Fig. $\ref{fig5}$. These values are slightly poorer than the typical energy resolutions of a single 2"$\times$2" LaBr$_3$(Ce) scintillator ($\approx$2.5$\%$ at 1 MeV \cite{CIEMALA200976,QUARATI2007115,seabury2006response}) and a single 3"$\times$3" NaI(Tl) scintillator (4.87$\%$ at 1173 keV \cite{DEMIR20213759,MOSZYNSKI2002259}), but are in good agreement with previous measurements with a single phoswich detector of PARIS, see Fig. 10 of Ref. \cite{Ghosh2016}. This observed degradation in energy resolution might arise from optical-coupling losses between the two scintillators and the PMT, the large-volume configurations, and imperfect gain matching among the phoswich detectors. In Fig. $\ref{fig5}$, the energy resolution of NaI(Tl) is characterized by $^{207}$Bi and $^{60}$Co sources \cite{gamma-source} while that of LaBr$_3$(Ce) was measured using $\beta$-delayed $\gamma$ rays of $^{80}$Ge \cite{PhysRevC.111.034303}. The data used to plot in Fig. \ref{fig5} are tabulated in Tab. \ref{tab2_1}, where the uncertainty in any number is given in parentheses after the number itself, e.g., 466.7(3) means 466.7 $\pm$ 0.3.

\begin{table}[htbp!] 
\centering
\fontsize{7.6pt}{10.83pt}\selectfont 
\caption{Data of Fig. \ref{fig5}. Energy resolutions of the LaBr$_3$(Ce) and NaI(Tl) crystals of PARIS are listed in normal (left four columns) and italic texts (right two columns), respectively. Note that the two values listed in the FWHM($\%$) columns correspond to the energy resolution of the LaBr$_3$(Ce) component obtained from the non-add-back and add-back within-one-cluster spectra, respectively. The uncertainty in any number is given in parentheses after the number itself, e.g., 466.7(3) means 466.7 $\pm$ 0.3. The energy resolution for NaI(Tl) is characterized by $^{207}$Bi and $^{60}$Co sources \cite{gamma-source} while for LaBr$_3$(Ce) by $\beta$-delayed $\gamma$ rays of $^{80}$Ge \cite{PhysRevC.111.034303}.}  \vspace{0.5mm}
\label{tab2_1}
\renewcommand\arraystretch{0.9}
\setlength{\tabcolsep}{3.0pt}
\begin{tabular}{lllllll}
$E_{\gamma}$ (keV) & FWHM ($\%$) & $E_{\gamma}$ (keV) & FWHM ($\%$) & $E_{\gamma}$ (keV) & FWHM ($\%$) \\ 
\hline
\hline
466.7(3)	&7.0(4) 6.9(3) & 1680.5(3) &3.4(1) 3.4(2) & $\it{569.698(2)}$ & $\it{10.8(1)}$ \\
520.0(3)	&6.2(4) 5.8(3) & 1772.6(4)	&2.9(4) 3.3(2) & $\it{1063.656(3)}$ & $\it{8.54(5)}$ \\
571.1(3)	&5.8(4) 5.0(2) & 2115.0(4)	&2.8(5) 2.0(5) & $\it{1173.228(3)}$ & $\it{7.86(6)}$ \\
809.1(3)	&5.3(7) 5.3(7) & 2600.2(5)	&3.6(7) 2.9(10) & $\it{1332.501(5)}$ & $\it{6.72(3)}$ \\
834.6(3)	&5.1(2) 4.8(3) & 2750.4(5)	&3.6(5) 3.5(5)\\
989.8(3)	&5.4(3) 3.8(5) & 2948.2(5)	&2.8(4) 2.6(2)\\
1005.2(3)	&4.8(5) 4.0(4) & 3108.9(5)	&2.7(9) 1.8(4)\\
1244.8(3)	&4.2(3) 4.2(4) & 3336.0(6)	&2.6(4) 2.4(5)\\
1306.7(3)	&5.1(3) 4.6(3) & 3664.6(6)	&3.1(3) 2.4(1)\\
1585.3(3)	&3.4(2) 2.7(7) & 5387.4(8)	&1.7(2) 1.7(2)\\

\hline
\end{tabular}
\end{table}

\section{Data sorting modes and detection efficiencies}

\begin{table*}[htbp]
\centering
\fontsize{8.63pt}{11pt}\selectfont 
\caption{Different data sorting modes of PARIS array}  \vspace{0.3mm}
\label{tab3}
\renewcommand\arraystretch{1.2}
\begin{tabular}{ll}
\hline
\hline
Modes 							& Descriptions \\
\hline
1	& Only LaBr$_3$(Ce) working individually (vetoed by NaI(Tl)) \\
2	& Only LaBr$_3$(Ce) (vetoed by NaI(Tl)) but add-back inside one cluster (nine phoswich detectors), add-back between LaBr$_3$(Ce) only 			\\
3	& Only LaBr$_3$(Ce) (vetoed by NaI(Tl)) but add-back totally (twenty-seven phoswiches of three clusters), add-back between LaBr$_3$(Ce) only			\\
4	& Summed spectrum from individual LaBr$_3$(Ce) (vetoed by NaI(Tl)) and individual NaI(Tl) \\
5	& Add back the signals from LaBr$_3$(Ce) and NaI(Tl) inside one phoswich (consider the cross-talk signals between two crystal)		\\
6	& Same as 5 but inside one cluster (consider the cross-talk signals between nine phoswiches) \\
\hline
\hline
\end{tabular}
\end{table*}

The PARIS data could be arranged into six sorting modes. For each mode, it is necessary to determine the individual detection efficiency. There is one Monte Carlo simulation package for PARIS, SToGS; see the user guide \cite{SToGS} to install and run the program. These six possible modes are listed in Tab. \ref{tab3}: 1-6. 1. Only LaBr$_3$(Ce) signals and each phoswich works individually, i.e. no add-back procedure, and NaI(Tl) crystals are used as veto detectors; 2. Only LaBr$_3$(Ce) vetoed by NaI(Tl) but being added back inside one cluster of nine phoswich detectors; 3. Only LaBr$_3$(Ce) vetoed by NaI(Tl) but being added back totally, i.e. within twenty-seven phoswich detectors of three clusters; 4. Summed spectrum from individual LaBr$_3$ (Ce) vetoed by NaI(Tl) meaning that the full energy is deposited in LaBr$_3$(Ce) and individual NaI(Tl); 5. Add back the signals from LaBr$_3$ and NaI(Tl) crystals inside one phoswich considering the cross-talk signals between two crystals, the part marked with "E, F" in Fig. $\ref{fig1}$; 6. Add back the signals from LaBr$_3$ and NaI(Tl) crystals inside one cluster, considering the cross-talk signal inside one cluster including nine LaBr$_3$(Ce) and nine NaI(Tl) crystals. It is important to highlight that Tab. \ref{tab3} delineates the different modes for sorting the data collected by the phoswich array, whereas Tab. \ref{tab1} lists the energy deposition scenarios arising when a $\gamma$ ray interacts with the array. For instance, the application of mode 2 results in an enhanced detection efficiency of the full-energy peaks corresponding to the situation B. Nevertheless, events from situations A are also present in the spectrum at the full-energy peaks. The NaI(Tl) veto was implemented by gating on the first peak illustrated in Fig. \ref{fig1}.

\begin{figure}[!htb]
\centering
\includegraphics[width=0.965\columnwidth]{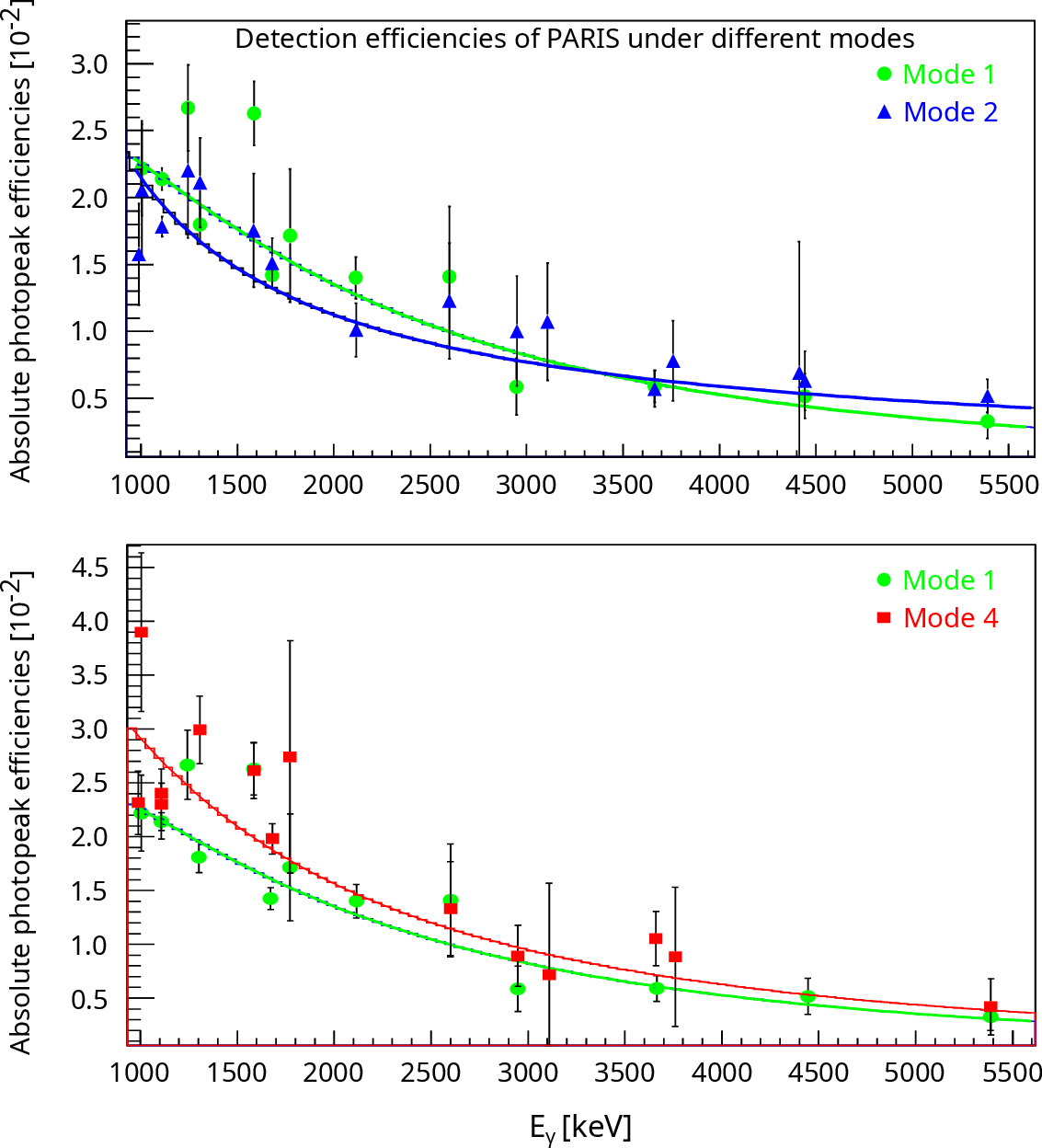}
\caption{PARIS detection efficiencies of three clusters (27 phoswich detectors) under different modes of analysis (see text), each one is compared with mode 1. The fit function used is from radware framework \cite{radware}, as shown in formula \ref{equ1}. The data presented here are from $\beta$-delayed $\gamma$ rays of $^{80}$Ge \cite{PhysRevC.111.034303}. The numbers of emitted $\gamma$ rays were obtained by combining the clover-HPGe-measured intensities with the GEANT4-simulated efficiency (see Fig. \ref{fig7}).}
\label{fig6}
\end{figure}

\begin{figure}[!htb]
\centering
\includegraphics[width=1.0\columnwidth]{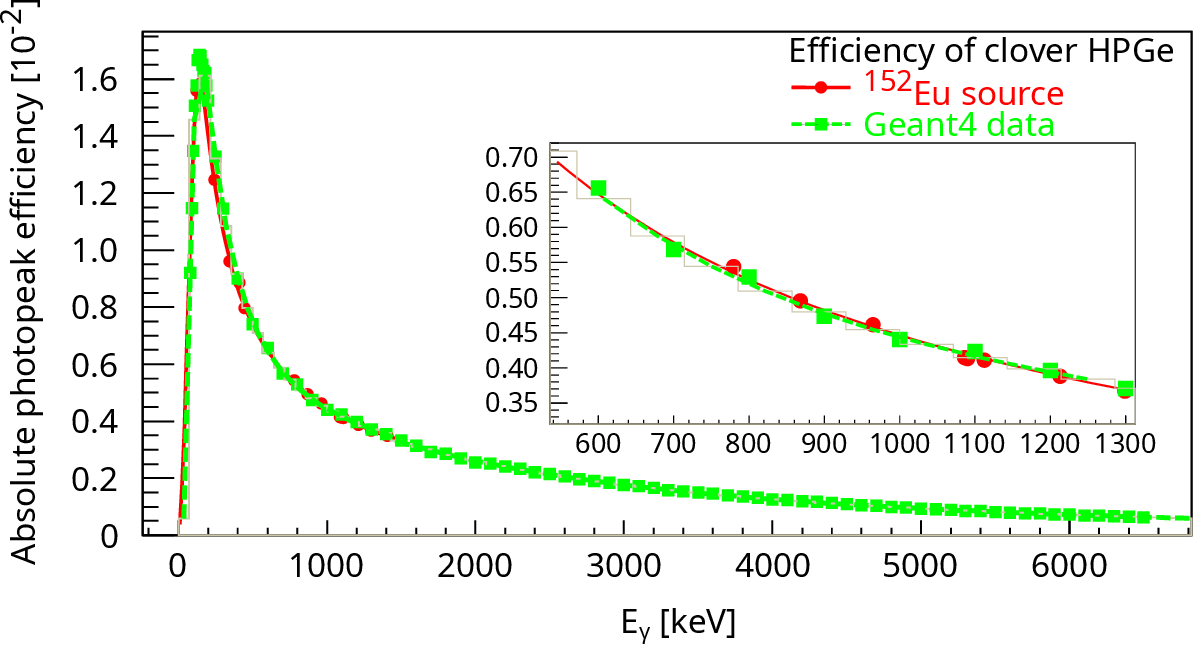}
\caption{Absolute full-energy peak efficiency of clover, added efficiency of 4 crystals: red points and the fitting curve (red solid line) are the present results from $^{152}$Eu source; green points and fitting curve (green dashed line) show efficiency obtained from Geant4 data. Fit function used is from radware framework \cite{radware}, as shown in formula \ref{equ2}. The insert is an enlarged figure of the region of 550-1300 keV to show the consistency of source data and simulation data. The Eurisys clover detector \cite{DUCHENE199990} is composed of four coaxial n-type Ge diodes with diameters of 50 mm and lengths of 70 mm. The distance between the end-cup of the clover detector and the source collection point is 74 mm.}
\label{fig7}
\end{figure}

\begin{figure}[!htb]
\centering
\includegraphics[width=1.0\columnwidth]{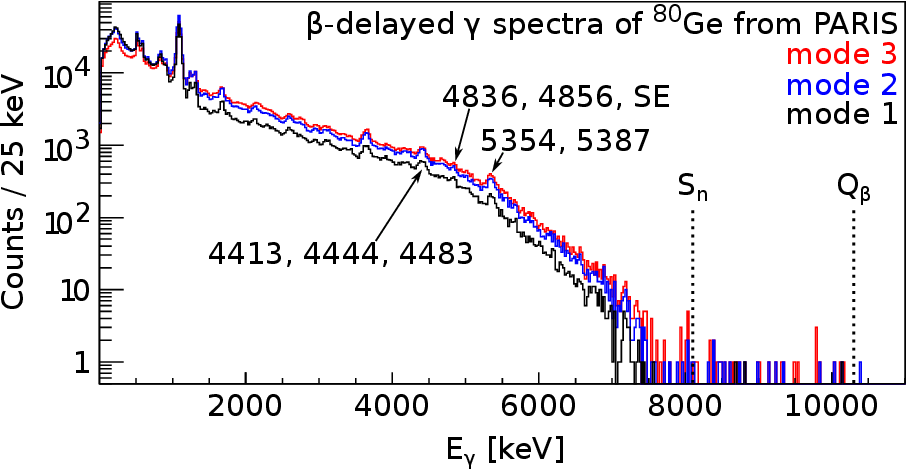}
\caption{$\beta$-delayed $\gamma$ spectra of $^{80}$Ge, obtained from PARIS in coincidence with the 659 keV $\gamma$-line (2$_1^+\rightarrow$0$_{g.s.}^+$ transition in $^{80}$Ge) detected in HPGe, black in mode 1 (non-add-back), blue in mode 2 {(add-back within one cluster)} and red in mode 3 {(addback with all phoswiches)}. Note that the numbers represent the energies of full-energy peaks in keV, while SE refers to the single-escape peaks of the 5354 and 5387 keV $\gamma$ rays, at 4843 and 4876 keV, respectively.}
\label{fig8}
\end{figure}

Fig. $\ref{fig6}$ shows the $\gamma$-ray detection efficiencies for three of the above sorting modes, which were obtained through normalizing them to the detection efficiency of the clover detector. The data presented are from $\beta$-delayed $\gamma$ rays of $^{80}$Ge \cite{PhysRevC.111.034303}. The numbers of emitted $\gamma$ rays were determined by combining the clover-HPGe-measured intensities with the GEANT4-simulated efficiency (see Fig. \ref{fig7}). The fit function with four parameters, shown in formula \ref{equ1}, is taken from radware framework \cite{radware}. The Eurisys clover detector \cite{DUCHENE199990} is composed of four coaxial n-type Ge diodes with diameters of 50 mm and lengths of 70 mm. The distance between the end-cup of the clover detector and the source collection point is 74 mm. The added absolute full-energy peak efficiency of four crystals is used to characterize the detection efficiencies of PARIS including three clusters (27 phoswich detectors) in different modes. Compared with mode 1, the full-energy peak detection efficiency is lower in the low-energy region in mode 2 while higher in the higher-energy region, above 3.5 MeV. In principle, add-back sums the energies recorded simultaneously in multiple detector segments when a single $\gamma$ ray deposits its energy in more than one segment due to Compton scattering and $\gamma$-escape events. In practice, however, [e.g., mode 2 (add-back within-one-cluster) used in the present work] pseudo-add-back occurs when more than one $\gamma$ ray from a $\beta$-delayed cascade hits a single cluster. This pseudo-add-back is particularly significant for low-energy $\gamma$ rays, which are located at the bottom of the decay level scheme and therefore have more coincidences with other $\gamma$ lines. Consequently, compared to mode 1, mode 2 exhibits a lower full-energy peak detection efficiency in the low-energy region [Fig. \ref{fig6}], and mode 3 similarly shows lower counts in the low-energy region, as illustrated in Fig. \ref{fig8}. In mode 4, i.e. summing events which deposit full energy in LaBr$_3$(Ce) (the first strong peak in Fig. \ref{fig1}) and events which only deposit full/partial energy in NaI(Tl) crystal (the second peak in Fig. \ref{fig1}), the full-energy peak detection efficiency is higher than in mode 1 in the whole energy region. 

\begin{equation} \label{equ1} 	
eff = D * e^{(A + B*x + C*x^2)},
\end{equation}
where x = log(E$_{\gamma}$/E1) and E1 = 1000 keV \cite{radware}.

For the detection efficiency of the clover HPGe, the low-energy region was calibrated using $^{152}$Eu source, as shown by the red curve in Fig. \ref{fig7}. For the high-energy region, e.g., up to 10 MeV, there is no physical calibration source. We performed numerical simulations \cite{geant4,haj2016method}, first reproducing the data obtained at lower energy with the $^{152}$Eu source then extending the efficiency curve to the high-energy region using same parameters. Good agreement can be found between the data collected from $^{152}$Eu source and those generated from Geant4 simulation, as shown in the inset of Fig. \ref{fig7}. Fig. \ref{fig7} shows the efficiency as a function of $\gamma$-ray energy of the clover HPGe and the associated fits. The fit function with seven parameters, shown in formula \ref{equ2}, is taken from radware framework \cite{radware}.

\begin{equation} \label{equ2} 
 eff = e^{[(A + B*x + C*x^2)^{-G} + (D + E*y + F*y^2)^{-G}]^{-1/G}},
\end{equation}
where x = log(E$_{\gamma}$/E1) and y = log(E$_{\gamma}$/E2), E1 = 100 keV; E2 = 1000 keV \cite{radware}.

For getting insights into the performance of the different modes directly, we present $\beta$-delayed $\gamma$ spectra of $^{80}$Ge in Fig. \ref{fig8}, obtained from PARIS in coincidence with the 659 keV $\gamma$-line (2$_1^+\rightarrow$0$_{g.s.}^+$ transition in $^{80}$Ge) in HPGe. One finds the statistics of the full-energy peaks, such as that of the 5387 keV $\gamma$-line, increase from mode 1 to mode 2 and from mode 2 to mode 3 while the escape peaks are suppressed. It is worth noting that the escape peaks and the Compton scattering background are already much weaker than the full-energy peaks in mode 1 because of the vetoing function of outer-volume NaI(Tl) crystals.

The role of the PARIS detectors usually is to provide the major contribution to the $\gamma$-ray efficiency in the highest part of the spectrum, above 6 MeV. In order to increase this efficiency, one should aim at the most compact geometrical positioning of the PARIS detectors. However, one still has to discriminate the $\gamma$ and neutron signals by their TOF \cite{Dey_2016wcd,PARIS_nresponse2018}. For that reason, a large distance such as 120 mm between the source and the detector end-cap is needed, which gives an 8.5 ns flight time difference between 1 MeV neutrons and $\gamma$ rays. Under this configuration, neutron-induced contamination in the $\beta$-delayed $\gamma$ spectrum of $^{80}$Ge obtained from PARIS can be effectively eliminated by applying a gate to select events with a TOF less than 4 ns.

As a final point, we investigate the role of multiplicity conditions in suppressing background and escape events for PARIS. Multiplicity refers to the number of $\gamma$ rays emitted during the deexcitation process, which is a cascade of transitions from the excited state to the ground state. For example, in the cascade of the 6046.6 keV state $\rightarrow$ 2$_1^+$ $\rightarrow$ 0$_{g.s.}^+$, two $\gamma$ rays are emitted with energies of 5387 keV and 659 keV. Therefore, in principle, the value of multiplicity is 2.

However, due to electron-positron pair production followed by positron annihilation, there are strong SE peaks (losing one 511 keV $\gamma$ ray) and DE peaks (losing two 511 keV $\gamma$ rays), especially in the high-energy region above 3 MeV. We found that this phenomenon is more pronounced in HPGe detectors, possibly because of the use of clover HPGe detectors, whereas it is not significant for PARIS even in mode 1 due to the vetoing function of the outer-volume NaI(Tl) crystals.

\begin{figure}[!htb]
\centering
\includegraphics[width=1.0\columnwidth]{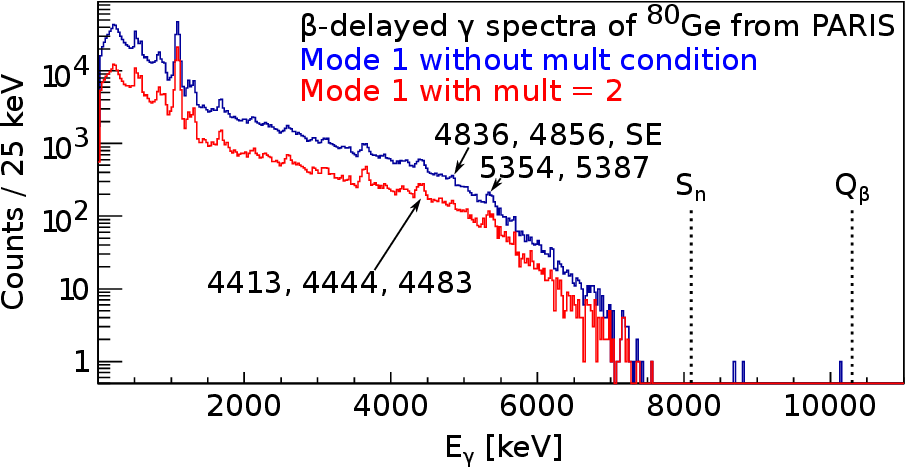}
\caption{The performance of multiplicity condition in suppressing escape events and background in PARIS detectors. $\beta$-delayed $\gamma$ spectra of $^{80}$Ge, obtained from PARIS in coincidence with 659 keV $\gamma$-line detected in HPGe, blue in mode 1 without multiplicity condition while red in mode 1 with multiplicity condition. Notably, "Mult" value is the number of $\gamma$ rays recorded by all detectors following a $\beta$-decay event. In $\gamma$-$\gamma$ coincidence analysis, multiplicities of events are equal to or greater than two. The spectra shown here are NaI(Tl) vetoed. Note that the numbers represent the energies of full-energy peaks in keV, while SE refers to the single-escape peaks of the 5354 and 5387 keV $\gamma$ rays, at 4843 and 4876 keV, respectively.}
\label{fig10}
\end{figure}

Additionally, in $\gamma$-$\gamma$ coincidence analysis, only events with multiplicities equal to or greater than two are selected. Note that applying the multiplicity condition results in some loss of statistics for the full-energy peaks as well. However, this technique is very useful to unambiguously distinguish escape peaks. Furthermore, comparing data obtained without the multiplicity condition to those with different multiplicity settings helps determine the multiplicity of the newly identified cascade.

Fig. \ref{fig10} presents results of imposing multiplicity conditions in PARIS. It is evident that multiplicity gating produces a slight effect on escape suppression, as demonstrated by the SEs of 5354 and 5387 keV, and leads to a modest enhancement of the signal-to-noise ratio. Nevertheless, there is a loss of statistics for full-energy peaks when applying multiplicity conditions. Notably, the spectra presented in Fig. \ref{fig10} are NaI(Tl) vetoed.

It should be emphasized that the spectra in Figs. \ref{fig8} and \ref{fig10} are free of pile-up contributions, as events with a Q$_{long}$-Q$_{short}$ ratio (Fig. \ref{fig1}) greater than 3 were excluded during energy calibration. The negligible, constant background above S$_n$ and the absence of background above Q$_{\beta}$ confirm this.

\section{Conclusion}

In conclusion, we investigated multiple data-sorting modes of the PARIS phoswich detector array for high-energy $\gamma$-ray detection, presenting results for time resolution, energy resolution, and detection efficiencies under different modes. Compared with a 2"$\times$2"$\times$2" LaBr$_3$(Ce) detector \cite{CIEMALA200976}, PARIS more effectively rejects escape events via the veto function of its outer-volume NaI(Tl) crystals in both individual and add-back modes, and rejects pile-up events by selecting events with Q$_{long}$/Q$_{short}$ smaller than 3. In contrast to the common approach of adding back the energies in LaBr$_3$(Ce) and NaI(Tl) to increase the peak-to-total ratio, using NaI(Tl) as a veto shield provides a superior trade-off for applications where spectral purity is essential. Multiplicity conditions provide additional suppression of escape events or background but simultaneously reduces the statistics of full-energy peaks. Notably, the methods presented in this work refer specifically to the $\beta$-decay experiment of $^{80g+m}$Ga conducted with three PARIS clusters comprising 27 phoswich detectors, rather than to a general report on the PARIS array or its overall performance. The developed method opens an avenue for further studies of the performance of the phoswich detector array in different data-sorting methods, which have nuclear experimental and $\gamma$-ray detecting satellite implications.

\section*{Acknowledgements}

The authors thankfully acknowledge the work of the ALTO technical staff for the excellent operation of the ISOL source. R.L. acknowledges support by the China Scholarship Council under Grant No.201804910509. C.D., A.K., and L.A.A. received funding from European Union's Horizon 2020 research and innovation program under Grant Agreement No. 771036 (ERC CoG MAIDEN). Use of the PARIS modular array from the PARIS Collaboration and Ge detectors from the French-UK IN2P3-STFC Gamma Loan Pool is acknowledged.

\bibliography{data_sorting_modes}
\end{document}